\documentclass[12pt]{article}
\pdfoutput=1

\usepackage{color}
\usepackage{amssymb,amsmath,bm,bbm}
\usepackage{epsf}
\usepackage{epsfig}
\usepackage{afterpage}
\usepackage{longtable}
\usepackage{booktabs}
\usepackage{caption}
\usepackage[dvipsnames]{xcolor}
\usepackage[linktoc=page,bookmarks=false,colorlinks=false,
linkbordercolor=RoyalBlue,citebordercolor=ForestGreen,urlbordercolor=CornflowerBlue]{hyperref}
\usepackage{latexsym,mathrsfs,dsfont}
\usepackage[normalem]{ulem} 
\usepackage[compress]{cite}
\usepackage{graphicx}
\usepackage{url}
\usepackage{paralist}
\usepackage{bbold}

\renewcommand{\arraystretch}{1.25}

\usepackage{fancyhdr}
\advance \headheight by 15.0truept       

\def \refeq#1{(\ref{#1})}
\def \refsec#1{Section~\ref{#1}}

\def \refapp#1{Appendix~\ref{#1}}
\def \reffig#1{Figure~\ref{#1}}
\def \reftab#1{Table~\ref{#1}}

\definecolor{darkgreen}{rgb}{0.0,0.6,0.0}

\DeclareMathOperator{\re}{Re}
\DeclareMathOperator{\im}{Im}

\newcommand{\ord}{\mathcal{O}}

\newcommand{\gev}{\, \text{GeV}}
\newcommand{\mev}{\, \text{MeV}}
\newcommand{\be}{\begin{equation}}
\newcommand{\ee}{\end{equation}}
\newcommand{\bsi}{B_6^{(1/2)}}
\newcommand{\bei}{B_8^{(3/2)}}

\def\epe{\varepsilon'/\varepsilon}

\def\kpn{K^+\rightarrow\pi^+\nu\bar\nu}
\def\klpn{K_L\rightarrow\pi^0\nu\bar\nu}

\newcommand{\OmHatEff}{\widehat\Omega_\text{eff}}

\newcommand{\alS}{\alpha_s}
\newcommand{\alE}{\alpha_\text{em}}

\newcommand \oL[1]{{\overline{#1}}}

\newcommand \lArr{\leftarrow}

\begin{document}

\vspace{-14mm}
\begin{flushright}
    {AJB-19-1}
\end{flushright}

\vspace{6mm}

\begin{center}
{\Large\bf
\boldmath{On the Importance of NNLO QCD and Isospin-breaking Corrections in $\epe$}
}
\\[12mm]
{\bf \large
  Jason Aebischer${}^a$,
  Christoph Bobeth${}^b$ and
  Andrzej~J.~Buras${}^c$
}
\\[0.8cm]
{\small
${}^a$Excellence Cluster Universe, TU M\"unchen, Boltzmannstr.~2, 85748~Garching, Germany \\[2mm]
      jason.aebischer@tum.de \\[2mm]
${}^b$Physik Department, TU M\"unchen, James-Franck-Stra{\ss}e, 85748~Garching, Germany \\[2mm]
      christoph.bobeth@tum.de\\[2mm]
${}^c$TUM Institute for Advanced Study, Lichtenbergstr.~2a, 85748~Garching, Germany \\[2mm]
      aburas@ph.tum.de
}
\end{center}

\vspace{6mm}

\abstract{
  \noindent
  Following the 1999 analysis of Gambino, Haisch and one of us, we stress that
  all the recent NLO analyses of $\epe$ in the Standard Model (SM) suffer from
  the {\em renormalization scheme} dependence present in the electroweak penguin
  contributions as well as from scale uncertainties in them related
  to the matching scale $\mu_W$ and in particular to $\mu_t$ in $m_t(\mu_t)$.
  We also reemphasize the important role of isospin-breaking  and QED effects
  in the evaluation of $\epe$. Omitting all these effects, as done in the 2015
  analysis   by RBC-UKQCD collaboration, and choosing as an example the QCD penguin
  ($Q_6$) and electroweak penguin ($Q_8$) parameters $\bsi$ and $\bei$ to be
  $\bsi = 0.80 \pm 0.08$ and $\bei = 0.76 \pm 0.04$ at $\mu = m_c=1.3\,\gev$,
  we find $(\epe)_\text{SM} = (9.4 \pm 3.5) \times 10^{-4}$, whereas including
  them results in $(\epe)_\text{SM} = (5.6\pm 2.4)\times 10^{-4}$.
  This is an example of an anomaly at the $3.3\,\sigma$ level, which would be
  missed without these corrections. NNLO QCD contributions to QCD penguins are
  expected to further enhance this anomaly. We provide a table for $\epe$ for
  different values of $\bsi$ and the isospin-breaking parameter $\OmHatEff$,
  that should facilitate monitoring the values of $\epe$ in the SM when the
  RBC-UKQCD calculations of hadronic matrix elements including isospin-breaking
  corrections and QED effects will improve with time.
}

\setcounter{page}{0}
\thispagestyle{empty}
\newpage



%
%
%

\section{Introduction}

The direct CP-violation in $K\to\pi\pi$ decays, represented by the ratio $\epe$,
plays a very important role in the tests of the Standard Model (SM) and more
recently in the tests of its possible extensions \cite{Buras:2018wmb}. In the
SM $\epe$ is governed by QCD penguins (QCDP) but receives also an important
contribution from the electroweak penguins (EWP), pointed out already in 1989
\cite{Flynn:1989iu, Buchalla:1989we}, that entering $\epe$ with the opposite
sign to QCDP suppress this ratio significantly. The partial cancellation of
these two contributions in addition to the significant uncertainties in
the evaluation of the hadronic matrix elements of QCDP and EWP operators is the
reason why until today a precise prediction for $\epe$ in the SM is still
missing.

The situation of $\epe$ in the SM by the end of 2017  could be briefly
summarized as follows:
\begin{itemize}
\item The analysis of $\epe$ by the RBC-UKQCD lattice QCD (LQCD) collaboration
  based on their 2015 results for $K\to \pi\pi$ matrix elements \cite{Bai:2015nea,
    Blum:2015ywa}, as well as the analyses performed in \cite{Buras:2015yba,
    Kitahara:2016nld} that are based on the same matrix elements but also
  include isospin breaking effects \cite{Cirigliano:2003nn, Cirigliano:2003gt},
  found $\epe$ in the ballpark of $(1-2) \times 10^{-4}$. This is by one order
  of magnitude below the experimental world average from NA48 \cite{Batley:2002gn}
  and KTeV \cite{AlaviHarati:2002ye, Worcester:2009qt} collaborations,
  \begin{align}
    \label{EXP}
    (\epe)_\text{exp} &
    = (16.6 \pm 2.3) \times 10^{-4} \,.
  \end{align}
  However, with an error in the ballpark of $5 \times 10^{-4}$ obtained in
  these analyses, one could talk about an $\epe$ anomaly of at most~$3\,\sigma$.
  Simultaneously, we note that the 2015 RBC-UKQCD result for the
  $\pi\pi$-strong-interaction phase $\delta_0$ of the isospin $I = 0$ amplitude
  is in almost $3\,\sigma$ conflict with the result from extrapolations in the
  chiral limit\cite{Colangelo:2001df}. This suggests that there were methodical
  problems with the 2015 RBC-UKQCD calculation, which were meanwhile successfully
  addressed, as will be reported later. As a conclusion, one has to be aware
  that for $I = 0$ the $K\to\pi\pi$ matrix elements, represented mainly by
  the parameter $B_6^{(1/2)}$, and hence also the 2015
  RBC-UKQCD result for $\epe$ suffer from an unaccounted systematic uncertainty.
\item An independent analysis based on hadronic matrix elements from the Dual
  QCD (DQCD) approach \cite{Buras:2015xba, Buras:2016fys} gave a strong support
  to the 2015 RBC-UKQCD result and moreover provided an \textit{upper bound} on
  $\epe$ in the ballpark of $6\times 10^{-4}$. However, in this approach
  the treatment of $\pi\pi$ strong interaction phases is presently problematic.
\item Chiral perturbation theory (ChPT) together with large-$N$
  considerations\footnote{See \cite{ Buras:2016fys, Buras:2018ozh} for a
  critical analysis of this approach as used in the context of $\epe$.} are
  used in \cite{Gisbert:2017vvj}, leading to a SM prediction of
  $\epe = (15 \pm 7) \times 10^{-4}$. The uncertainties are larger than in \cite{Buras:2015yba, Kitahara:2016nld}, reflecting in part the difficulties
  in matching long-distance and short-distance contributions in this framework,
  but are also of parametric origin due to low-energy constants.
  Consequently, the predicted central value is one order of magnitude larger
  compared to DQCD and lattice results of 2015, but with a small tension
  of $1.6\,\sigma$ in view of the large uncertainties.
\end{itemize}

Recently progress towards an improved estimate of $\epe$ in the SM has been made:
\begin{itemize}
\item The RBC-UKQCD collaboration is expected to present this year new values
  of the $K\to\pi\pi$ hadronic matrix elements, most importantly the
  parameter $B_6^{(1/2)}$. In particular the discrepancy in
  the prediction of the $\pi\pi$-strong-interaction phase $\delta_0$ has been
  identified \cite{Wang:2019nes, Kelly:2019yxg, Christ:2019kaon, Kelly:2019wfj}
  in the form of excited-state contamination requiring the introduction of
  additional $\pi\pi$ operators in the simultaneous fits to allow for a better
  isolation of the ground state. It can be expected that the statistical errors
  will decrease, though less dramatically as assumed before due to the additional
  operators. Unfortunately, the inclusion of isospin-breaking and QED effects
  will still take more time.
\item An improved estimate of isospin-breaking corrections to $\epe$ has been
  presented in \cite{Cirigliano:2019cpi} increasing moderately the role of
  these corrections in suppressing $\epe$. The updated ChPT analysis
  \cite{Cirigliano:2019cpi} resulted in the value
  \begin{align}
    \label{Pich}
    (\epe)_\text{ChPT} &
    = (14 \pm 5) \times 10^{-4} \,, \qquad (\text{ChPT}-2019),
  \end{align}
  in full agreement with the experimental world average \refeq{EXP}.
\item The preliminary result on NNLO QCD corrections to QCDP contributions
  \cite{Cerda-Sevilla:2016yzo, Cerda-Sevilla:2018hjk} demonstrates
  significant reduction of various scale uncertainties, foremost of $\mu_c$,
  and indicates an additional modest suppression of $\epe$.
\end{itemize}

In contrast to the expected RBC-UKQCD result, the ChPT analysis includes
isospin-breaking and QED corrections but the known difficulties in matching
long-distance and short-distance contributions in this approach imply a large
uncertainty. In particular, the absence of the so-called meson evolution in
ChPT that suppresses $\epe$ within the DQCD approach \cite{Buras:2016fys, Buras:2018ozh}
is responsible for the poor matching and according to the latter authors
responsible in part for the large value of $\epe$ in \refeq{Pich}.
The DQCD analysis in \cite{Buras:2018lgu} demonstrates through the example of
BSM matrix elements in $K^0-\oL{K}^0$ mixing that the effects of meson evolution
are included in the present LQCD calculations. As shown in \cite{Buras:2018lgu},
neglecting this evolution in the case of $K^0-\oL{K}^0$ mixing would miss the
values of the relevant hadronic matrix elements by factors of $2-4$, totally misrepresenting
their values obtained by three LQCD collaborations \cite{Carrasco:2015pra,
  Jang:2015sla, Garron:2016mva, Boyle:2017skn, Boyle:2017ssm}. The
fact that in $K^0-\oL{K}^0$ mixing the FSI are absent allows to study the impact
of meson evolution better than it is possible in $K\to\pi\pi$ decays. Yet, as
demonstrated in \cite{Buras:2015xba, Buras:2016fys} these important QCD dynamics
must also be present in $K\to\pi\pi$ and is also required by the proper matching
of long-distance and short-distance contributions. Therefore it is expected to
suppress the result for $\epe$ in~\refeq{Pich}.

Now all the 2015 analyses of $\epe$ and the one in \cite{Cirigliano:2019cpi}
used the known Wilson coefficients at the NLO level \cite{Buras:1991jm,
Buras:1992tc, Buras:1992zv, Buras:1993dy, Ciuchini:1992tj, Ciuchini:1993vr} in
the naive dimensional regularization (NDR) scheme \cite{Buras:1989xd}. But
already in \cite{Buras:1999st} and recently in \cite{Aebischer:2018csl, Buras:2018ozh}
it has been pointed out that without NNLO QCD corrections to the EWP contribution the
results for $\epe$ are renormalization-scheme dependent and exhibit significant
non-physical dependences on the scale $\mu_t$ at which the top-quark mass
$m_t(\mu_t)$ is evaluated as well as on the matching scale $\mu_W$.

Fortunately, all these uncertainties have been significantly reduced in the NNLO
matching at the electroweak scale performed in \cite{Buras:1999st} and it is
of interest to look at them again in the context of new analyses with the goal
to improve the present estimate of $\epe$. Additional importance in such an
analysis is the finding in \cite{Buras:1999st} that these corrections further
suppress $\epe$ relative to the NLO results performed in the NDR scheme.

In view of the fact that LQCD calculations contain both the meson
evolution and FSI, the optimal strategy for the evaluation of $\epe$ as of
2019 appears to be as follows:
\begin{enumerate}
\item Use future RBC-UKQCD results for hadronic matrix elements of the
  dominant QCDP ($Q_6$) and EWP ($Q_8$) operators, represented by the parameters
  $\bsi$ and $\bei$ respectively -- with improved values of $\pi\pi$-strong-interaction
  phases $\delta_{0,2}$ -- but determine hadronic matrix elements
  of $(V-A)\otimes(V-A)$ operators from the experimental data on the real
  parts of the $K\to\pi\pi$ amplitudes as done in \cite{Buras:1993dy,Buras:2015yba}.
\item Use the result for isospin-breaking and QED corrections from
  \cite{Cirigliano:2019cpi}, which are compatible with the ones obtained
  already 30 years ago in \cite{Buras:1987wc}.
\item Use the NNLO QCD contributions to EWP in \cite{Buras:1999st} in order
  to reduce the unphysical renormalization scheme and scale dependences.
\item Include NNLO QCD contributions to QCDP from \cite{Cerda-Sevilla:2016yzo,
  Cerda-Sevilla:2018hjk} in order to reduce left-over renormalization scale
  uncertainties.
\end{enumerate}
In view of the fact that meson evolution and the remaining three
effects tend to suppress $\epe$, whereas the aforementioned systematic
uncertainties in the 2015 RBC-UKQCD calculation of $B_6^{(1/2)}$ related to
FSI effects discussed before in connection with $\delta_0$ could tend to
increase it, as does FSI in the case of ChPT, it could well happen that future
LQCD predictions of $\epe$ in the SM increase only moderately to end up in
or a bit above  the ballpark $\epe\approx (5\pm2) \times 10^{-4}$ of the
expectation based on the DQCD approach in \cite{Buras:2018ozh}.

The main goal of our paper is to illustrate how a future result from RBC-UKQCD
would be affected by the inclusion of known isospin-breaking and QED corrections
from~\cite{Cirigliano:2019cpi} in point 2. and the NNLO QCD contributions to
EWP in \cite{Buras:1999st} in point 3. We also comment on the expected size of
NNLO QCD contributions to QCDP from \cite{Cerda-Sevilla:2016yzo, Cerda-Sevilla:2018hjk}
in point~4. leaving a detailed analysis of them to these authors.

Our paper is organized as follows. In \refsec{sec:2} we recall a number of basic
formulae that will be used in the rest of our paper. In \refsec{sec:3} we address
the issue of scale and renormalization scheme dependences at the NLO level resulting
from the EWP sector. We illustrate the size of these unphysical effects present
at the NLO level that  would increase the errors quoted in the existing
NLO analyses but are significantly reduced when NNLO QCD corrections to EWP
contributions from \cite{Buras:1999st} are taken into account. To this end we use
first as an example particular values for the hadronic parameters $\bsi$ and $\bei$
quoted in the abstract of our paper and quantify also the role of
isospin-breaking and QED corrections from \cite{Cirigliano:2019cpi}. As the
values of $\bsi$ and the size of isospin-breaking corrections are expected to
dominate the theoretical uncertainties in $\epe$ for some time, we present in
\refsec{sec:numerics} a table of the SM values of $\epe$ for different
$\bsi$ and the isospin-breaking parameter $\OmHatEff$ that should facilitate
monitoring the SM estimates of $\epe$ when the LQCD calculations of hadronic
matrix elements including isospin-breaking corrections and QED effects will
improve with time. A brief summary and an outlook are given in \refsec{sec:summary}.

%
%
%
\section{Basic Formulae}
\label{sec:2}

%
%
\subsection{An analytic formula}

As in \cite{Buras:2015yba}, our starting expression is the formula
\begin{align}
  \label{eprime}
  \frac{\varepsilon'}{\varepsilon} &
  = -\,\frac{\omega_+}{\sqrt{2}\,|\varepsilon_K|}
  \left[\, \frac{\im A_0}{\re A_0}\, (1 - \OmHatEff)
       -   \frac{1}{a} \, \frac{\im A_2}{\re A_2} \,\right],
\end{align}
where \cite{Cirigliano:2003gt, Cirigliano:2019cpi}
\begin{align}
  \label{OM+}
  \omega_+ & = a\,\frac{\re A_2}{\re A_0} = (4.53 \pm 0.02) \times 10^{-2}, &
  a & = 1.017, &
  \OmHatEff & = (17.0\pm 9.1) \times 10^{-2} \,.
\end{align}
Here $a$ and $\OmHatEff$ summarize isospin-breaking corrections and include
strong isospin violation $(m_u \neq m_d)$, the correction to the isospin limit
coming from $\Delta I=5/2$ transitions and electromagnetic corrections as first
summarized in \cite{Cirigliano:2003nn, Cirigliano:2003gt} and recently updated
in \cite{Cirigliano:2019cpi}. Our $\OmHatEff$, defined by
\begin{align}
  \OmHatEff &
  \equiv \Omega_\text{IB} - \Delta_0|_{\alpha = 0} - f_{5/2} \,,
\end{align}
differs from $\Omega_\text{eff}$ in \cite{Cirigliano:2003nn, Cirigliano:2003gt,
Cirigliano:2019cpi}
as in contrast to these papers it does not include EWP contributions to
$\im A_0$, summarized in these papers by $\Delta_0$. This is indicated
here by $\Delta_0|_{\alpha = 0}$, which contains the remaining contributions
only. We find it more natural to
calculate $\im A_0$ including both QCD and EWP contributions as this allows to
keep track of NP contributions to $\im A_0$.  The dominant EWP contribution to
$\epe$ is of course present in $\im A_2.$ In fact the RBC-UKQCD collaboration
includes EWP contributions to $\im A_0$ as well. We note also that the latest
central value for $\Omega_\text{IB} = 0.25 \pm 0.08$ from \cite{Cirigliano:2019cpi}
agrees perfectly with the one obtained already 30 years ago in \cite{Buras:1987wc}.

\begin{table}
\centering
\renewcommand{\arraystretch}{1.3}
\begin{tabular}{|l|rrrr|}
\hline
             & $a^\text{QCDP}$ & $a_6^{(1/2)}$ & $a^\text{EWP}$ & $a_8^{(3/2)}$ \\
\hline\hline
  \multicolumn{5}{|c|}{$\mu_W = \mu_t = m_W$} \\
\hline
  NLO        & $-4.19$       & $17.68$       & $-2.08$       & $8.25$        \\
  NNLO (EWP) & $-4.19$       & $17.68$       & $-2.00$       & $8.82$        \\
\hline
  \multicolumn{5}{|c|}{$\mu_W = m_W$ and  $\mu_t = m_t$} \\
\hline
  NLO        & $-4.18$       & $17.63$       & $-1.94$       & $7.22$        \\
  NNLO (EWP) & $-4.18$       & $17.63$       & $-2.03$       & $8.51$        \\
\hline
\end{tabular}
\caption{\small
  Coefficients entering the semi-numerical formula of \refeq{eq:semi-num-1}.}
  \label{tab:semi-num-1}
\end{table}

The real parts of the isospin amplitudes $A_{0,2}$ in \refeq{eprime} are then
extracted from the branching ratios on $K\to\pi\pi$ decays in the isospin limit.
In the limit $a = 1$ and $\OmHatEff = 0$ the formula in \refeq{eprime} reduces
to the one used in RBC-UKQCD \cite{Bai:2015nea, Blum:2015ywa}, where all
isospin-breaking corrections except for EWP contributions at the NLO level have
been set to zero.

Using the technology in \cite{Buras:2015yba} we arrive at the
formula
\begin{align}
  \label{eq:semi-num-1}
  \frac{\varepsilon'}{\varepsilon} & =
  \im \lambda_t \cdot \left[ 
    a (1 - \OmHatEff) \left(a^\text{QCDP} + a_6^{(1/2)} \bsi \right)
    - a^\text{EWP} - a_8^{(3/2)} \bei
  \right]\,,
\end{align}
with the numerical values of the coefficients given in
\reftab{tab:semi-num-1} at NLO and NNLO from EWPs as discussed below.
Explicit formulae for 
$a^\text{QCDP} = a_0^{(1/2)} - b\, a_{0,\text{EWP}}^{(1/2)}$, 
$a^\text{EWP} = a_0^{(3/2)} - a_{0,\text{EWP}}^{(1/2)}$,
$a_6^{(1/2)}$ and $a_8^{(3/2)}$ in terms of Wilson coefficients and 
$\re A_{0,2}$ are given in \cite{Buras:2015yba}, where we have 
introduced $a_{0,\text{EWP}}^{(1/2)}$ as the EWP contribution to $a_0^{(1/2)}$
and $b^{-1} = a (1 - \OmHatEff)$. The values of 
the Wilson coefficients used by us are collected in \refapp{app:wilson-coeffs},
whereas $\lambda_t = V_{td}^{} V^*_{ts}$ is the relevant CKM combination.

%
%
\subsection{\boldmath The parameters $\bsi$ and $\bei$}

The $\bsi$ and $\bei$ parameters, that enter the formula \refeq{eq:semi-num-1},
are defined as follows
\begin{align}
  \label{eq:Q60}
  \langle Q_6(\mu) \rangle_0 &
  = -\,4 h \left[\frac{m_K^2}{m_s(\mu) + m_d(\mu)}\right]^2 (F_K - F_\pi) \,\bsi \,
  = -0.473\, h \bsi \gev^3 \,,
\\
  \label{eq:Q82}
  \langle Q_8(\mu) \rangle_2 &
  = \sqrt{2} h \left[ \frac{m_K^2}{m_s(\mu) + m_d(\mu)} \right]^2 F_\pi \,\bei\,
  = 0.862\, h \bei \gev^3 \,,
\end{align}
with \cite{Buras:1985yx,Buras:1987wc}
\begin{align}
  \label{LN}
  \bsi & = \bei = 1\,,
\end{align}
in the large-$N$ limit.
The dimensionful parameters entering \refeq{eq:Q60}, \refeq{eq:Q82} have been
calculated at $\mu = m_c$ using \cite{Aoki:2016frl}
\begin{align}
  \label{FpFK}
  m_K & = 497.614 \mev , &
  F_\pi & = 130.41(20) \mev , &
  \frac{F_K}{F_\pi} & = 1.194(5) \,,
\end{align}
\begin{align}
  \label{msmd}
  m_s(m_c) & = 109.1(2.8) \mev , &
  m_d(m_c) & = 5.44(19) \mev \,.
\end{align}
We have introduced the factor $h$ in order to emphasize different normalizations
of these matrix elements present in the literature. For instance RBC-UKQCD and
\cite{Buras:2015yba} use $h=\sqrt{3/2}$, while \cite{Buras:2015xba, Buras:2016fys,
Gisbert:2017vvj, Cirigliano:2019cpi} use $h = 1$.

As an example we will first use the values
\begin{align}
  \label{Lbsi}
  \bsi(m_c) & = 0.80\pm 0.08 , &
  \bei(m_c) & = 0.76\pm 0.04 ,
\end{align}
to be compared with the 2015 values $\bsi(m_c) = 0.57 \pm 0.19$ and
$\bei(m_c) = 0.76 \pm 0.05$ from RBC-UKQCD \cite{Bai:2015nea, Blum:2015ywa}.

While we do not expect significant modification of the RBC-UKQCD result for
$\bei$ through the improvements on FSI, taking the arguments on the impact of
FSI on $\bsi$ from ChPT into account \cite{Cirigliano:2019cpi}, we allow
for an enhancement of $\bsi$, which however is still consistent with the arguments
in \cite{Buras:2016fys} that the suppression of $\bsi$ by meson evolution below
unity is stronger than its enhancement by FSI. We emphasize that the choice of
$\bsi$ in \refeq{Lbsi} is only an example. Other examples will be presented in
Section~\ref{sec:numerics}, where also values of $\bsi>1.0$, in the spirit of
\cite{Cirigliano:2019cpi}, are considered. We anticipate a significant reduction
of the error on $\bsi$ in the new results of RBC-UKQCD collaboration relative
to its 2015 analysis so that the expectations from \cite{Buras:2016fys} and
\cite{Cirigliano:2019cpi} will be tested.

%
%
%
\section{Scale uncertainties at NLO}
\label{sec:3}

It should be emphasized that although the NLO QCD analyses of
$\epe$ in \cite{Buras:1991jm, Buras:1992tc, Buras:1992zv, Buras:1993dy,
  Ciuchini:1992tj, Ciuchini:1993vr} reduced renormalization scheme dependence in
the QCDP sector, the dependence of $\epe$ on the choice of $\mu_t$ in
$m_t(\mu_t)$ remained. This dependence can only be removed through the NNLO QCD
calculations, but in the QCDP sector it is already weak at the NLO level because
of the weak dependence of the QCDP contributions on $m_t$. On the other hand, as
pointed out already in \cite{Buras:1999st}, the EWP contributions at the NLO
level suffer from a number of unphysical dependences.
\begin{itemize}
\item First of all there is the renormalization-scheme dependence with $\epe$ in
  the HV scheme, as used in \cite{Ciuchini:1992tj, Ciuchini:1993vr}, generally
  smaller than in the NDR scheme used in \cite{Buras:1991jm, Buras:1992tc,
    Buras:1992zv, Buras:1993dy}. In what follows we will consider only the NDR
  scheme as this is the scheme used by the RBC-UKQCD collaboration and other
  analyses listed above.
\item The dependence on $\mu_t$, which is much larger than in the QCDP sector
  because the EWP contributions exhibit much stronger dependence on $m_t$.
  Increasing $\mu_t$ makes the value of $m_t$ smaller, decreasing the EWP
  contribution and thereby making $\epe$ larger. At NLO there is no QCD
  correction that could cancel this effect.
\item The dependence on the choice of the matching scale $\mu_W$. It turns out
  that with increasing $\mu_W$ in the EWP contribution, the value of $\epe$
  decreases.
\end{itemize}

\begin{figure}
\centering
  \includegraphics[width=0.5\textwidth]{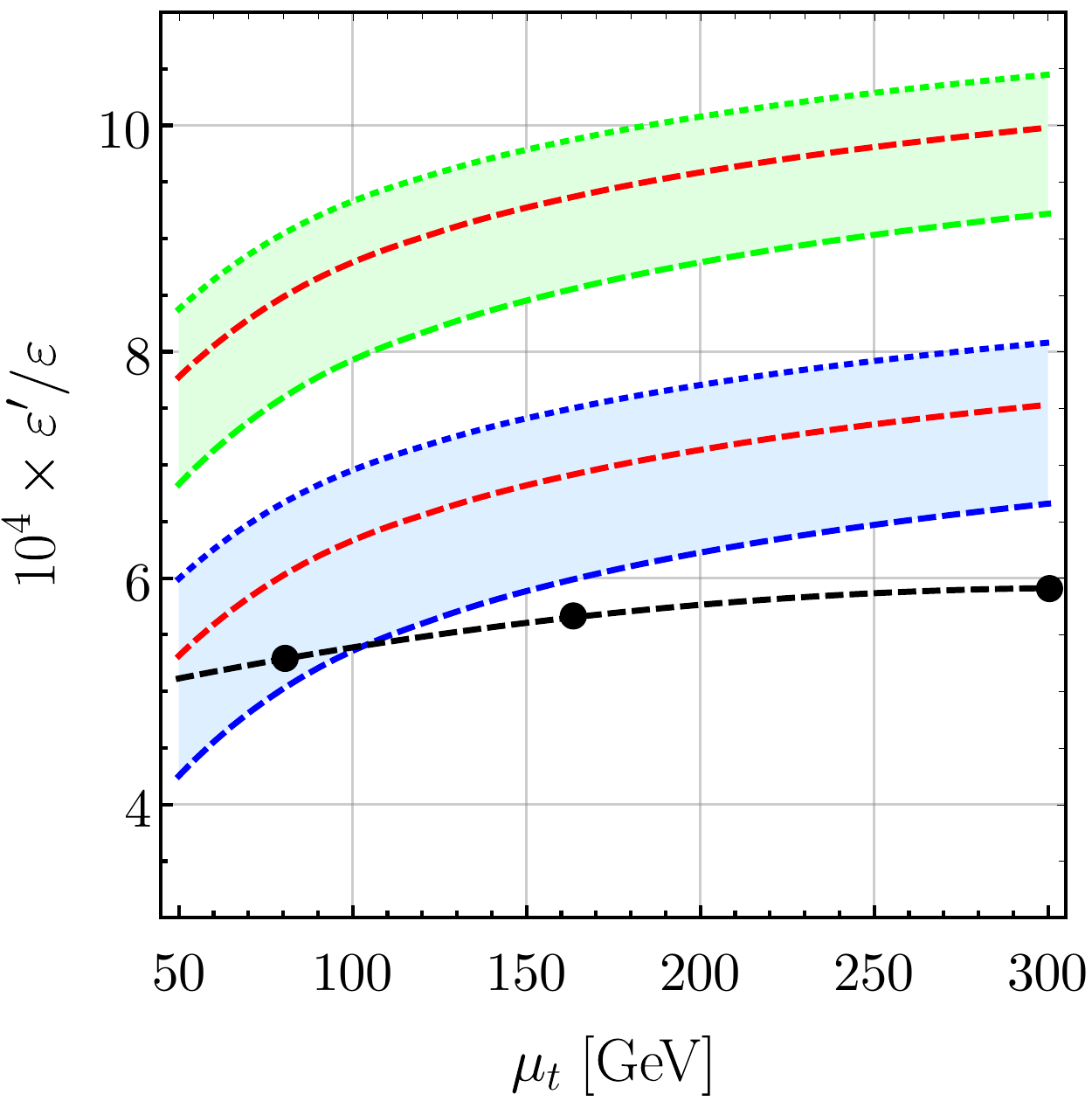}
\caption{\small The dependence of $\epe$
    for $\bsi = 0.80$ on the scale $\mu_t$  of $m_t(\mu_t)$ for three values of
    the matching scale $\mu_W = \{60,\, 80,\, 120\}\,$GeV [dotted, red, dashed]
    for $\OmHatEff = 0.0$ [green] and $\OmHatEff = 0.17$ [blue]. The black dots
    show the NNLO result for $\OmHatEff = 0.17$ at these scales $\mu_t$
    from \cite{Buras:1999st} with interpolation shown by the dashed line.
    We set $\bei=0.76$.}
\label{fig:epe-muW}
\end{figure}

One should note that the scales $\mu_W$ and $\mu_t$ can be chosen to be equal
or different and they could be varied independently in the ranges
illustrated in \reffig{fig:epe-muW} implying
significant uncertainties in the NLO prediction for $\epe$ as demonstrated in
\cite{Buras:1999st}. In obtaining the values in \reftab{tab:semi-num-1} we
provide the two settings from \cite{Buras:1999st}: $i)$ $\mu_W = \mu_t = m_W$
as well as $ii)$ $\mu_W = m_W$ and $\mu_t = m_t$. For example $ii)$ has been
used in \cite{Buras:2015yba}. Other choices of these scales would significantly
change the NLO values of $\epe$ with significantly reduced change  when NNLO
corrections to EWPs are included.

We next evaluate $\epe$ for the values of $\bsi$ and $\bei$
given in \refeq{Lbsi} and
\begin{itemize}
\item set $\mu_W = m_W$ and $\mu_t = m_t$ in the NLO formulae in the NDR scheme,
\item set $\OmHatEff = 0.0$,
\end{itemize}
as done by RBC-UKQCD. This results at NLO in
\begin{align}
  \label{RBCUKQCD}
  (\epe)_{\text{NLO},\, \OmHatEff = 0.0} &
  = (9.4 \pm 3.5) \times 10^{-4} \,,
\end{align}
that is  a value by a factor of 7 larger than the 2015 result from the
RBC-UKQCD collaboration. The quoted error is a guess estimate based on the
uncertainties in \refeq{Lbsi} and scale uncertainties as well as omission
of isospin-breaking effects ignoring the known signs of these effects. But as
we will see soon its precise size is irrelevant for the point we want to make.
The result in \refeq{RBCUKQCD}  is compatible  with experiment \refeq{EXP}
with a tension of $1.7\,\sigma$.

At first sight it would appear that this result confirms the claims in
\cite{Gisbert:2017vvj} and \cite{Cirigliano:2019cpi} as~\refeq{RBCUKQCD}
is quite consistent with \refeq{Pich}. But such a conclusion would be false
as we will illustrate now.

Indeed as stated above at the NLO level significant dependences on $\mu_W$ and
$\mu_t$ are present and the impact of a non-vanishing $\OmHatEff$ is very significant.
In order to exhibit these dependences we vary in \reffig{fig:epe-muW} the matching
scale $\mu_W$ independently of the scale $\mu_t$ at which the top-quark mass
$m_t(\mu_t)$ is evaluated and plot $\epe$ versus $\mu_t$ for the three values of
$\mu_W = \{60,\, 80,\, 120\}\,$GeV. We show these dependences both for
$\OmHatEff = 0.0$ [green] and $\OmHatEff = 0.17$ [blue]. They are very significant.

Fortunately all these uncertainties have been significantly reduced in the NNLO
matching at the electroweak scale performed in \cite{Buras:1999st}. In the
NDR scheme, used in all recent analyses, these corrections enhance for
$i)$ $\mu_W = \mu_t = m_W$ the EWP contribution by roughly $7\,\%$ and
for $ii)$ $\mu_W = m_W$ and $\mu_t=m_t$ by $16\,\%$. Thereby they
imply a {\em negative} shift in $\epe$ that depends on $\bei$ and $\im\lambda_t$
and in fact, as just stated and evident from \reffig{fig:epe-muW} on the chosen values of
$\mu_t$ and $\mu_W$ in the NLO expressions. Including NNLO QCD corrections
in question and using $\OmHatEff$ in \refeq{OM+} the result in \refeq{RBCUKQCD}
is changed to
\begin{align}
  \label{ABB}
  \epe  & \, = (5.6 \pm 2.4) \times 10^{-4} \,,
\end{align}
which compared with the experimental value in \refeq{EXP} signals an anomaly
at the level of $3.3\,\sigma$. In \reftab{tab:epsp} below, we have set
$\im \lambda_t = 1.4 \times 10^{-4}$. For the result in \refeq{RBCUKQCD}
and \refeq{ABB} we have
used $\im \lambda_t = (1.43\pm 0.04) \times 10^{-4}$,
based on recent analyses of the unitarity triangle by the bayesians
(``UTfitter") and frequentists (``CKMfitter") that can be found in
\cite{Bona:2006ah} and \cite{Charles:2015gya}, respectively.

The error budget, discussed in \refapp{app:errors} and summarized in
\reftab{tab:errors-BP}, would imply the parametric theoretical error of
$2.3\times 10^{-4}$.
We increased it in order to take into account left-over scale uncertainties both
in the EWP sector discussed here and in the QCDP sector analyzed at NNLO
in \cite{Cerda-Sevilla:2016yzo, Cerda-Sevilla:2018hjk}. But one should keep in
mind that the central value in \refeq{ABB} will be shifted down by NNLO QCD
corrections to QCDP by about $0.5\times 10^{-4}$ as indicated in  the preliminary
plots in \cite{Cerda-Sevilla:2016yzo, Cerda-Sevilla:2018hjk} without modifying
the error in (\ref{ABB}). We are looking forward to the final results of these
authors.

Our NNLO central value in \refeq{ABB}, represented in \reffig{fig:epe-muW}
by the black points at $\mu_t = \{m_W,\, m_t,\, 300\,\gev\}$, is much less
dependent on $\mu_t$. This exercise shows also the importance of isospin-breaking
corrections. They are  significantly larger than the NNLO QCD corrections to EWP
contributions.

It should be emphasized that in \cite{Buras:1999st} complete
$\ord(\alpha_W \alpha_s)$ and $\ord(\alpha_W\alpha_s\sin^2\theta_W m_t^2)$
corrections, with $\alpha_W = \alpha/\sin^2\theta_W$, to the Wilson coefficients
$C_{7-10}(\mu)$ of EWP operators at $\mu = m_c$ have been calculated. In
particular as demonstrated in Section~3 of that paper no three-loop anomalous
dimensions of involved operators are necessary to find these corrections. See
formula (3.14) of that paper. In order to complete the NNLO analysis of EW
contributions one should calculate $m_t$-independent
$\ord(\alpha_W\alpha_s\sin^2\theta_W)$ corrections, which as argued in
\cite{Buras:1999st} are much smaller than the ones included here.

Much more difficult is the NNLO analysis of QCD penguin contributions which in
addition to two-loop calculations requires three-loop anomalous dimension
matrices \cite{Cerda-Sevilla:2016yzo, Cerda-Sevilla:2018hjk}, obtained
fortunately already in \cite{Gorbahn:2004my}.

Inspection of the formulae in \cite{Buras:2015yba} together with the numbers in
\reftab{tab:semi-num-1} shows that the NNLO matching corrections lead mainly to
an enhancement of the coefficient $a_8^{(3/2)} \propto y_8(\mu)$ of $\bei$ by
$1.07\, (1.16)$ due to $y_8(\mu)$, whereas the NNLO impact on $\epe$ through $y_{7,9,10}$
is negligible due to the smaller matrix elements multiplying them. The size of
the enhancement depends on the choice of the matching scale $\mu_W$ and the
$\mu_t$ scale in $m_t(\mu_t)$ in the NLO expressions. The implications of these
uncertainties for $\epe$ are clearly seen in \reffig{fig:epe-muW}.

It should be added that the shift of including NNLO corrections in question on
$\epe$ is independent of the value of $\bsi$, but its magnitude depends of
course on $\im \lambda_t$ and $\bei$ that we set to $0.76$ in
\reffig{fig:epe-muW}. It reads for the choice $\mu_W = m_W$
\begin{align}
  \label{eq:epe_NNLO-shift}
  \Delta\epe|_\text{NNLO'} &
  = -\frac{\im \lambda_t}{1.4\cdot 10^{-4}} \times
    \left\{ \begin{array}{rl}
            0.80 \; \bei & \qquad \mu_t = m_W \\
            1.81 \; \bei & \qquad \mu_t = m_t
            \end{array} \right. \,.
\end{align}
For other choices of $\mu_W$ and $\mu_t$ the shifts can be read off from
\reffig{fig:epe-muW}. The prime in~\refeq{eq:epe_NNLO-shift} reminds us that
small $\ord(\alpha_W\alpha_s\sin^2\theta_W)$ NNLO corrections have not been
included in this formula.

In contrast to \cite{Buras:2015yba}, in obtaining the result for $\epe$ in
\refeq{ABB} we anticipated that in a future analysis of $\epe$ the LQCD values
of matrix elements $\langle Q_6(m_c) \rangle_0$ and $\langle Q_8(m_c) \rangle_2$
and not the parameters $\bsi(m_c)$ and $\bei(m_c)$ will be used,
which avoids the uncertainties in~\refeq{FpFK} and \refeq{msmd} that enter
the extraction of $\bsi$ and $\bei$. Therefore, when calculating the error in
\refeq{ABB} the uncertainties in \refeq{FpFK} and \refeq{msmd} have not been
included. Then the values of the matrix elements in question corresponding to
$\bsi(m_c)$ and $\bei(m_c)$ in \refeq{Lbsi} are given as follows
\begin{align}
  \label{LQCDQ6Q8}
  \langle Q_6(m_c) \rangle_0 &
  = h\,(-0.38\pm 0.04)\gev^3 \,,
&
  \langle Q_8(m_c) \rangle_2 &
  = h\,(0.66 \pm 0.03)\gev^3 \,.
\end{align}
The error budget is discussed in \refapp{app:errors} and summarized in
\reftab{tab:errors-BP}.

What is still missing are NNLO QCD corrections to QCDPs which on the
basis of \cite{Cerda-Sevilla:2016yzo, Cerda-Sevilla:2018hjk} are expected to
further suppress $\epe$, albeit the effect appears to be smaller than the one of
NNLO QCD contributions to EWPs. One could in principle question the inclusion of
the latter contributions while leaving out NNLO corrections to QCDPs.
Yet these two different NNLO contributions do not have anything to do with each
other. In particular while NLO QCD corrections to QCDPs remove already
some scale and renormalization scheme dependences present at the LO, in the EWP
sector these unphysical scheme dependences are first removed at the NNLO level
\cite{Buras:1999st}.

%
%
%
\section{Numerical analysis}
\label{sec:numerics}

\begin{table}
\centering
\renewcommand{\arraystretch}{1.4}
\begin{tabular}{|l|rrrrrrrrrrr|}
\hline
  $\bsi$
              &       0.50 &  0.55 &  0.60 & 0.65 & 0.70 & 0.75 & 0.80 &  0.85 &  0.90 &  0.95 &  1.00 \\
\hline\hline
  $\OmHatEff$ & \multicolumn{11}{|c|}{NLO} \\
\hline
  $0.0$ (A)   &       2.25 & 3.50  & 4.75  & 6.01 & 7.26 & 8.51 & 9.76 & 11.02 & 12.27 & 13.52 & 14.77 \\
  $0.0$ (B)   &       1.63 & 2.89  & 4.14  & 5.40 & 6.65 & 7.91 & 9.16 & 10.42 & 11.67 & 12.93 & 14.18 \\
  $0.0$ (C)   &       0.75 & 2.01  & 3.27  & 4.53 & 5.79 & 7.05 & 8.30 &  9.56 & 10.82 & 12.08 & 13.34 \\
\hline & \multicolumn{11}{|c|}{NNLO ($\mu_t = m_W$)} \\
\hline
  $0.0$       &       0.02 &  1.28 &  2.54 & 3.80 & 5.06 & 6.32 & 7.58 &  8.83 & 10.09 & 11.35 & 12.61 \\
  $0.10$      &      -0.64 &  0.50 &  1.63 & 2.76 & 3.89 & 5.03 & 6.16 &  7.29 &  8.42 &  9.56 & 10.69 \\
  $0.15$      &      -0.97 &  0.10 &  1.17 & 2.24 & 3.31 & 4.38 & 5.45 &  6.52 &  7.59 &  8.66 &  9.73 \\
  $0.20$      &      -1.30 & -0.29 &  0.71 & 1.72 & 2.73 & 3.74 & 4.74 &  5.75 &  6.76 &  7.76 &  8.77 \\
  $0.25$      &      -1.63 & -0.69 &  0.26 & 1.20 & 2.15 & 3.09 & 4.03 &  4.98 &  5.92 &  6.87 &  7.81 \\
  $0.30$      &      -1.96 & -1.08 & -0.20 & 0.68 & 1.56 & 2.44 & 3.33 &  4.21 &  5.09 &  5.97 &  6.84 \\
\hline
  \multicolumn{12}{c}{} \\
\hline
   $\bsi$
              &       1.05 &  1.10 &  1.15 &  1.20 &  1.25 &  1.30 &  1.35 &  1.40 &  1.45 &  1.50 &  1.55 \\
\hline\hline
  $\OmHatEff$ & \multicolumn{11}{|c|}{NLO} \\
\hline
  $0.0$ (A)   &      16.03 & 17.28 & 18.53 & 19.78 & 21.03 & 22.29 & 23.54 & 24.79 & 26.04 & 27.30 & 28.55 \\
  $0.0$ (B)   &      15.44 & 16.69 & 17.95 & 19.20 & 20.46 & 21.71 & 22.97 & 24.22 & 25.48 & 26.73 & 27.90 \\
  $0.0$ (C)   &      14.60 & 15.86 & 17.11 & 18.37 & 19.63 & 20.89 & 22.15 & 23.41 & 24.67 & 25.92 & 27.18 \\
\hline & \multicolumn{11}{|c|}{NNLO ($\mu_t = m_W$)} \\
\hline
  $0.0$       &      13.87 & 15.13 & 16.39 & 17.64 & 18.90 & 20.16 & 21.42 & 22.68 & 23.94 & 25.19 & 26.45 \\
  $0.10$      &      11.82 & 12.95 & 14.09 & 15.22 & 16.35 & 17.49 & 18.62 & 19.75 & 20.88 & 22.02 & 23.15 \\
  $0.15$      &      10.80 & 11.87 & 12.94 & 14.01 & 15.08 & 16.15 & 17.22 & 18.29 & 19.36 & 20.43 & 21.50 \\
  $0.20$      &       9.78 & 10.78 & 11.79 & 12.80 & 13.80 & 14.81 & 15.82 & 16.82 & 17.83 & 18.84 & 19.84 \\
  $0.25$      &       8.75 &  9.70 & 10.64 & 11.58 & 12.53 & 13.47 & 14.42 & 15.36 & 16.30 & 17.25 & 18.19 \\
  $0.30$      &       7.73 &  8.61 &  9.49 & 10.37 & 11.25 & 12.13 & 13.02 & 13.90 & 14.78 & 15.66 & 16.54 \\
\hline
\end{tabular}
\caption{\small
  The ratio $10^4 \times \epe$ at NNLO for different values of the isospin
  corrections $\OmHatEff$ and the parameter $\bsi(m_c)$ with more details in
  \refapp{app:wilson-coeffs} and fixed value of $\bei = 0.76$ and $\im \lambda_t
  = 1.4 \times 10^{-4}$. In the first three rows we provide for comparison
  the NLO result for $\mu_t = 300\,$GeV (A), $\mu_t = m_t$ (B) and $\mu_t = m_W$ (C),
  respectively.}
  \label{tab:epsp}
\end{table}

Our analysis shows that the largest remaining uncertainties in the evaluation of
$\epe$ are present in the values of $\langle Q_6(m_c)\rangle_0$ (or $\bsi$) and
$\OmHatEff$. In \reftab{tab:epsp} we give $\epe$ as a function of these two
parameters for $\bei=0.76$. This table should facilitate monitoring the values
of $\epe$ in the SM when the LQCD calculations of hadronic matrix elements
including isospin-breaking corrections and QED effects will improve with
time. We observe a large sensitivity of $\epe$ to $\bsi$, but for $\bsi\ge 0.7$
also the dependence on $\OmHatEff$ is significant.

Finally, it is of interest to ask how large a central value of $\bsi$ should be
in order to reproduce the central experimental value in \refeq{EXP}. It turns
out to be $\bsi = 1.40$, in total disagreement with \refeq{Lbsi}. The central
value in \refeq{Pich} is obtained for $\bsi = 1.24$.

%
%
%
\section{Summary and Outlook}
\label{sec:summary}

Our analysis and in particular the comparison of the results in \refeq{RBCUKQCD}
and \refeq{ABB} as well as the \reftab{tab:epsp} demonstrates the importance of
NNLO QCD corrections and of isospin-breaking effects. Anticipating that the new
RBC-UKQCD analysis will find $\bsi(m_c)< 1.0$ as hinted by DQCD, the values of
$\epe$ in the SM will be significantly below the data. Our example with $\bsi(m_c)$
in the ballpark of $0.80 \pm 0.08$ illustrates a significant anomaly in $\epe$ of
about $3.3\,\sigma$. If confirmed by new RBC-UKQCD analysis this would turn out
to be one of the largest anomalies in flavour physics present in a single observable and
comparable to the anomaly in the flavour conserving $(g-2)_\mu$. Moreover, this
would be presently the only significant anomaly as far as CP-violation is concerned
with the possible exception of the one present in $B\to \pi K$ decays \cite{Aaij:2018tfw},
as recently reviewed in \cite{Fleischer:2017vrb, Fleischer:2018bld, Faisel:2018bvs,
Datta:2019tuj}.

However, even if our expectations for $\epe$ in the SM would be confirmed by
new RBC-UKQCD results, in order to obtain a better assessment which NP is
responsible for this anomaly it is very important to perform a number of the
following steps:
\begin{itemize}
\item Obtain satisfactory precision on $\langle Q_6(m_c)\rangle_0$ or $\bsi$.
\item Reduce the error on $\OmHatEff$. In particular isospin-breaking and QED
  effects should be taken into account in LQCD calculations.
\item Even if the insight from DQCD allowed us to identify the dynamics (meson
  evolution) responsible for this anomaly, at least a second lattice QCD
  collaboration should calculate $K\to \pi\pi$ matrix elements and $\epe$.
\item Further reduce the short-distance uncertainties, in particular in the QCD
  penguin sector. But the subleading NNLO QCD contributions to the electroweak
  penguin sector should be also evaluated.
\item Calculation of BSM $K\to\pi\pi$ hadronic matrix elements of four-quark
  operators by lattice QCD that presently are known only in the DQCD
  \cite{Aebischer:2018rrz}.
\end{itemize}

There have been numerous BSM analyses of $\epe$ which we collect in
\reftab{eprimeanomaly}. Here we just mention that the leptoquark models, with
possible exception of the vector $U_1$ model, are not capable in explaining this
anomaly because of the constraints from rare Kaon decays
\cite{Bobeth:2017ecx}. This shows how crucial correlations of $\epe$ with other
observables in a given NP scenario are.  As indicated in \reftab{eprimeanomaly},
they have been analyzed in other NP scenarios. In particular, very recently,
a correlation of hinted anomalies in $\epe$ and $B\to\pi K$ decays has been
pointed out in the context of models with $U(2)^3$ flavour symmetry in
\cite{Crivellin:2019isj}.

Also the lessons gained from the SMEFT analysis in \cite{Aebischer:2018csl}
should be very helpful in identifying NP behind hinted $\epe$ anomaly. Such
a general analysis allows to take the constraints from other
processes, in particular from electroweak precision tests and collider processes
into account. To this end the master formula for $\epe$ \cite{Aebischer:2018quc}
valid in any extension of the SM should facilitate the search for the dynamics
behind the anomaly in question.

\begin{table}
\renewcommand{\arraystretch}{1.3}
\centering
\resizebox{\columnwidth}{!}{
\begin{tabular}{|c|c|c|}
\hline
  NP Scenario & References  & Correlations with
\\
\hline\hline
  LHT
& \cite{Blanke:2015wba}
& $\klpn$
\\
  $Z$-FCNC
& \cite{Buras:2015jaq, Bobeth:2017xry, Endo:2016tnu}
& $\kpn$ and $\klpn$
\\
  $Z^\prime$
& \cite{Buras:2015jaq}
& $\kpn$, $\klpn$ and $\Delta M_K$
\\
  Simplified Models
& \cite{Buras:2015yca}
& $\klpn$
\\
  331 Models
& \cite{Buras:2015kwd, Buras:2016dxz}
& $b\to s\ell^+\ell^-$
\\
  Vector-Like Quarks
& \cite{Bobeth:2016llm}
& $\kpn$, $\klpn$ and $\Delta M_K$
\\
  Supersymmetry
& \cite{Tanimoto:2016yfy, Kitahara:2016otd, Endo:2016aws, Crivellin:2017gks, Endo:2017ums}
& $\kpn$ and $\klpn$
\\
  2HDM
& \cite{Chen:2018ytc, Chen:2018vog}
& $\kpn$ and $\klpn$
\\
  Right-handed Currents
& \cite{Cirigliano:2016yhc, Alioli:2017ces}
& EDMs
\\
  Left-Right Symmetry
& \cite{Haba:2018byj, Haba:2018rzf}
&  EDMs
\\
  Leptoquarks
& \cite{Bobeth:2017ecx}
& all rare Kaon decays
\\
  SMEFT
& \cite{Aebischer:2018csl}
& several processes
\\
  $\text{SU(8)}$
& \cite{Matsuzaki:2018jui}
&  $b\to s\ell^+\ell^-$, $\kpn$, $\klpn$
\\
  Diquarks
& \cite{Chen:2018dfc, Chen:2018stt}
& $\varepsilon_K$, $\kpn$, $\klpn$
\\
  3HDM + $\nu_R$
& \cite{Marzo:2019ldg}
&  $R(K^{(*)})$, $R(D^{(*)})$
\\
  Vectorlike compositeness
& \cite{Matsuzaki:2019clv}
&  $R(K^{(*)})$, $R(D^{(*)})$, $\varepsilon_K$, $\kpn$, $\klpn$
\\
  $U(2)^3$ flavour symmetry
& \cite{Crivellin:2019isj}
& hadronic $B\to K\pi$, $B_{s,d}\to (KK, \pi\pi)$, $B_s\to \phi (\rho^0,\pi^0)$
\\
\hline
\end{tabular}
}
\caption{Papers studying implications of the $\epe$ anomaly.
  \label{eprimeanomaly}
}
\end{table}

%
%
%

\section*{Acknowledgments}

This research was done and financially  supported by the DFG cluster
of excellence ``Origin and Structure of the Universe''.

%
%
%
\appendix

%
%
%
\section{Wilson coefficients}
\label{app:wilson-coeffs}

Here we summarize the $\Delta S = 1$ Wilson coefficients at the scale
$\mu = m_c = 1.3\,$GeV in the NDR-$\oL{\text{MS}}$ scheme using the NLO RG
evolution from \cite{Buras:1993dy}.  The numerical input is fixed to values in
\reftab{tab:num-in-WC}.  The running of the couplings at the low-energy scale
are $\alS(m_c) = 0.3764$ and $1/\alE(m_c) = 133.84$. The threshold crossings are
at $\mu_4 = 4.2\,$GeV for $N_f = 5 \to 4$ and $\mu_3 = 1.3\,$GeV for
$N_f = 4 \to 3$ quark flavours.

We will use the results in \cite{Buras:1999st} to demonstrate the numerical
impact of the dominant NNLO matching corrections that resolve the NLO
renormalization scheme ambiguities for the two choices $\mu_t = m_W$ and
$\mu_t = m_t$. As given in Table 2 (Table 3) of \cite{Buras:1999st} they lead to a rescaling
of $y_{7, \ldots, 10}(\mu)$ at the low-energy scale\footnote{Note that in Table 2 of
\cite{Buras:1999st} the entry $C_8(\mu_K) = 0.142$ at NLO$_\text{NDR}$ disagrees
with Figure~6 $C_8(\mu_K) = 0.149$, where the latter will be used here.}
of about $1.07\,(0.92)$, $1.07\,(1.16)$, $0.89\,(0.98)$ and $0.76\,(0.85)$ for
$\mu_t = m_W\, (m_t)$ to the NNLO' values in \reftab{tab:wc-at-muc}, which we adapt
in the numerics. The prime in this Table indicates that still small $m_t$-independent
$\ord(\alpha_W\alpha_s\sin^2\theta_W)$ corrections are not included and
NNLO corrections to $y_{3,4,5,6}$ are neglected as well.

\begin{table}
\centering
\renewcommand{\arraystretch}{1.3}
\begin{tabular}{|lll|lll|}
\hline
  Parameter
& Value
& Ref.
&  Parameter
& Value
& Ref.
\\
\hline\hline
  $\alS^{(5)}(m_Z)$                & $0.1181(11)$                         & \cite{Tanabashi:2018oca}
& $m_Z$                            & $91.1876$     GeV                    & \cite{Tanabashi:2018oca}
\\
  $\alE^{(5)}(m_Z)$                & $1/127.955(10)$                      & \cite{Tanabashi:2018oca}
& $m_W$                            & $80.385$ GeV                         & \cite{Tanabashi:2018oca}
\\
  $s_W^2 = \sin^2(\theta_W)$       & $0.23126$                            & \cite{Tanabashi:2018oca}
& $m_t^\text{pole}$                & $173.1(6)$ GeV                       & \cite{Tanabashi:2018oca}
\\
\hline
\end{tabular}
\caption{\small
   Numerical input for Wilson coefficients.
}
  \label{tab:num-in-WC}
\end{table}

\begin{table}
\centering
\renewcommand{\arraystretch}{1.3}
\begin{tabular}{|c|rr|rr|}
\hline
& \multicolumn{2}{|c|}{$\mu_t = m_W$}
& \multicolumn{2}{|c|}{$\mu_t = m_t$}
\\
\hline
                & NLO        & NNLO'      & NLO           & NNLO'  \\
\hline\hline
  $z_1$         & $-0.394$  & $\lArr$   & $\lArr$   & $\lArr$  \\
  $z_2$         & $ 1.202$  & $\lArr$   & $\lArr$   & $\lArr$  \\
\hline
  $y_3$         & $ 0.027$  & $\lArr$   & $\lArr$   & $\lArr$  \\
  $y_4$         & $-0.055$  & $\lArr$   & $-0.054$  & $\lArr$  \\
  $y_5$         & $ 0.006$  & $\lArr$   & $\lArr$   & $\lArr$  \\
  $y_6$         & $-0.083$  & $\lArr$   & $-0.082$  & $\lArr$  \\
\hline
  $y_7/\alE$    & $-0.024$  & $-0.026$  & $-0.038$  & $-0.035$ \\
  $y_8/\alE$    & $ 0.131$  & $ 0.141$  & $ 0.119$  & $ 0.138$ \\
  $y_9/\alE$    & $-1.495$  & $-1.330$  & $-1.406$  & $-1.378$ \\
  $y_{10}/\alE$ & $ 0.533$  & $ 0.405$  & $ 0.497$  & $ 0.422$ \\
\hline
\end{tabular}
\caption{\small
  The $\Delta S = 1$ Wilson coefficients at $\mu = m_c = 1.3\,$GeV in the
  NDR-$\oL{\text{MS}}$ scheme for the renormalization scale $\mu_W = m_W$
  and $\mu_t = m_W$ or $\mu_t = m_t$ using NLO and partial NNLO matching results
  for $y_{7,\ldots,10}$. The symbol $\lArr$ indicates that there are now
  changes in the numerical value within the adapted approximation.
}
  \label{tab:wc-at-muc}
\end{table}

%
%
%
\section{Error budget}
\label{app:errors}

We summarize the error budget leading to the result in \refeq{ABB} in
\reftab{tab:errors-BP}. The scale uncertainties after the inclusion of
NNLO corrections to both QCDP and  EWP are not shown as they have negligible
impact on the final error.

\begin{table}
\centering
\renewcommand{\arraystretch}{1.3}
\begin{tabular}{|c|c|}
\hline
  Parameter     &  $10^4\times\delta(\epe)$   \\
\hline\hline
  $\bsi$        &  $\pm 1.7$       \\
  $p_3$         &  $\pm 0.5$       \\
  $\bei$        &  $\pm 0.5$       \\
  $p_5$         &  $\pm 0.6$       \\
  $q$           &  $\pm 0.1$       \\
  $B_8^{(1/2)}$ &  $\pm 0.2$       \\
  $p_{72}$      &  $\pm 0.05$      \\
  $p_{70}$      &  $\pm 0.05$      \\
\hline
  $m_t^\text{pole}$ & $\pm 0.05$   \\
  $\alS(m_Z)$   &  $\pm 0.1$       \\
  $\im \lambda_t$ & $\pm 0.1$      \\
\hline
  $\OmHatEff$   &  $\pm 1.3$       \\
\hline
\end{tabular}
\caption{\small
   Table of the absolute error of $\epe$ for benchmark point \refeq{Lbsi} with
   input parameters from \refeq{OM+} and~\refeq{Lbsi}. The absolute
   error of $\epe$ from these parametric uncertainties becomes $10^4 \times
   \delta(\epe) = 2.3$ when added in quadrature.
}
  \label{tab:errors-BP}
\end{table}

%
%
%

\renewcommand{\refname}{R\lowercase{eferences}}

\addcontentsline{toc}{section}{References}

\footnotesize

\bibliographystyle{JHEP}
\bibliography{Bookallrefs}

\end{document}